
\def\half {{1 \over 2}}
\def\dz{{\partial _z}}
\def\dzbar{{\partial _{\bar z}}}
\tolerance=5000

\overfullrule=0pt 
\baselineskip=18pt
\raggedbottom
\centerline{\bf A New Approach to the Green-Schwarz Superstring}
\vskip 12pt
\centerline{Nathan Berkovits}
\vskip 12pt
\centerline{Maths Dept., King's College, Strand, London, WC2R 2LS, United
Kingdom}
\vskip 12pt
\centerline{June 1993}
\vskip 12pt
\centerline{KCL-TH-93-9}
\vskip 24pt

\centerline{Based on talks given at SUSY '93 (Northeastern
University) and Strings '93 (Lawrence Berkeley Laboratory)}

\vskip 24pt
\centerline {\bf Abstract}

By replacing two of the bosonic scalar superfields of the N=2 string
with fermionic scalar superfields (which shifts $d_{critical}$ from
(2,2) to (9,1)), a quadratic action for the ten-dimensional
Green-Schwarz superstring is obtained. Using the usual N=2 super-Virasoro
ghosts, one can construct a BRST operator, picture-changing operators, and
covariant vertex operators for the Green-Schwarz superstring. Superstring
scattering amplitudes with an arbitrary number of loops and external
massless states are then calculated by evaluating correlation functions
of these vertex operators on N=2 super-Riemann surfaces, and integrating
over the N=2 super-moduli.

These multiloop superstring amplitudes have been proven to be SO(9,1)
super-Poincar\'e invariant (by constructing the super-Poincar\'e
generators and writing the amplitudes in manifest SO(9,1) notation),
unitary (by showing agreement with amplitudes obtained using the
light-cone gauge Green-Schwarz formalism), and finite (by explicitly
checking for divergences in the amplitudes when the Riemann surface
degenerates). There is no multiloop ambiguity in these
Green-Schwarz scattering amplitudes since spacetime-supersymmetry is
manifest (there is no sum over spin structures), and therefore the
moduli space can be compactified.
\vfil
\eject

There are two main reasons for studying the Green-Schwarz superstring.
The first reason is to obtain a more efficient method for calculating
superstring amplitudes than is possible using the NSR formalism. Because
of manifest spacetime-supersymmetry in the Green-Schwarz formalism,
there is no need to perform GSO projections or sum over spin structures.
As will be described later in this talk, this simplifies calculations
involving external fermions (there are no square-root cuts and fermionic
vertex operators do not require ghosts), removes the multiloop ambiguity
(the moduli space can be compactified since there is no need for a cutoff
associated with summing over spin structures), and allows a direct proof
of finiteness (the amplitudes can be explicitly checked to be free of
divergences).$^{1-4}$
Although these amplitude calculations are perturbative in
the string coupling constant, they may be useful for studying possible
quantum corrections to general relativity, or for finding new symmetries
in superstring theory.

The second reason for studying the Green-Schwarz superstring is to get
a better understanding of super-Yang-Mills and supergravity. Since
two-dimensional non-linear sigma models provide a natural framework for
studying the massless fields of the string, one would expect that by
constructing the appropriate sigma model for the Green-Schwarz superstring,
one could learn something about off-shell super-Yang-Mills and
supergravity. For the case of the four-dimensional Green-Schwarz superstring,
this expectation has been confirmed (four-dimensional super-Yang-Mills
and supergravity fields are scalar potentials and Kahler vectors of
N=2 non-linear sigma models)$^{5,6}$,
while for the ten-dimensional case, work is
still in progress. Hopefully, a better understanding of these massless
supersymmetric field theories will be useful in unraveling how superstring
theory produces a consistent quantum theory of gravity.

Until recently, the only method available for calculating Green-Schwarz
superstring amplitudes was the light-cone gauge method in which all
world-sheet symmetries, including conformal invariance, are
non-manifest.$^{7-9}$
Amplitudes are calculated in this method by evaluating correlation functions
of light-cone vertex operators and interaction-point operators on a
two-dimensional surface, and integrating over the positions of the
vertex-operator
punctures and the moduli of the surface. These interaction-point
operators are required for Lorentz invariance and can be understood as
the light-cone analog of picture-changing operators, which come from
integrating out the world-sheet gravitini.$^1$ However unlike picture-changing
operators, their locations on the surface are completely fixed, and in
fact are extremely complicated functions of the puncture positions, the
surface moduli, and the $P^+$ momenta of the external states. Because
of this complication, the light-cone Green-Schwarz method has not
yet produced manifestly Lorentz-invariant expressions for any amplitude
with more than one loop or more than four external states. An additional
problem of the light-cone Green-Schwarz method is that in order to remove
non-physical divergences when interaction-points coincide, one needs
to introduce contact-terms whose
precise form has not yet been determined.$^{10}$

Recently, a new method$^2$ has been developed for calculating Green-Schwarz
superstring amplitudes which starts from the manifestly N=(2,0) worldsheet
supersymmetric action:
$$S=\int d^2 z d^2 \kappa [X^a \dzbar X^{\bar a} +W^-\dzbar\Theta^+
-W^+\dzbar\Theta^-],\eqno(1)$$
subject to the chirality constraints, $D_+ X^{\bar a}=D_+ \Theta^-=
D_- X^a=D_-\Theta^+=0 $  ($a,\bar a$ range from 1 to $A$ and
$D_\pm =\partial_{\kappa^\pm}+\kappa^\mp \dz$), the N=2 superconformal
constraint,
$D_+ X^a D_- X^{\bar a}+D_- W^- D_+\Theta^+ +D_+ W^+ D_-\Theta^- =0$,
and the global constraint,
$D_+ \Theta^+ D_-\Theta^- -\half(\Theta^+\dz\Theta^- +\Theta^-\dz\Theta^+)
=\dz X^+$ for some real superfield $X^+$.
This action is manifestly invariant under an $SU(A)\times U(1)$ subset of the
super-Poincar\'e group, which includes the $2A+2$ spacetime-supersymmetry
transformations, $\delta X^a=\epsilon^a \Theta^+$, $\delta X^{\bar a}=
\epsilon^{\bar a}\Theta^-$, $\delta\Theta^+=\epsilon^+$, $\delta\Theta^-=
\epsilon^-$, $\delta W^-=\epsilon^a X^{\bar a}$, $\delta W^+=\epsilon^{\bar a}
X^a$. Although only the heterotic superstring (ignoring lattice degrees
of freedom) will be discussed in this talk, the new method easily
generalizes to non-heterotic Green-Schwarz superstrings. However up to now,
it is possible only for the heterotic case to obtain this quadratic action
by partially gauge-fixing a manifestly Lorentz-covariant action.$^{1,11}$

The action of equation 1 is just the usual N=2 string action,$^{12,13}$
except that the two ``longtitudinal'' pairs of bosonic scalar superfields,
$(X^0,X^{\bar 0})$ and $(X^d,X^{\bar d})$, have been exchanged for two pairs
of fermionic scalar superfields, $(\Theta^+,W^-)$ and $(\Theta^-,W^+)$. Note
that although the $W^\pm$ superfields are not chiral, only $D_+ W^+$ and
$D_- W^-$ contribute to the action. Since the central charge contribution
of the longtitudinal superfields is thereby flipped from $+6$ to $-6$,
the critical N=2 string now contains four pairs of transverse superfields,
$X^a=x^a+\kappa^+\Gamma^a$ and
$X^{\bar a}=x^{\bar a}+\kappa^-\Gamma^{\bar a}$ for $a,\bar a=1$ to 4, which
describe the usual light-cone Green-Schwarz content of eight scalar bosons
and eight spin-$\half$ fermions (because of spectral flow, these eight
spin-$\half$ fermions could alternatively be treated as four spin-0
and four spin-1 fermions).

The longtitudinal degrees of freedom of the Green-Schwarz superstring are
described by the four bosonic and four fermionic components of the
superfields, $\Theta^\pm=\theta^\pm+\kappa^\pm \lambda^\pm$ and
$D_\pm W^\pm=w^\pm+\kappa^\mp \varepsilon^\pm$, which are subject to the
global constraint,
$\lambda^+\lambda^- -\half(\theta^+\dz\theta^- +\theta^-\dz\theta^+)
=\dz x^+$ for some real field $x^+$.
Since this global constraint on $\lambda^\pm$
commutes with the N=2 stress-energy tensor,
it does not affect the conformal anomaly calculation. However in order
to construct vertex operators and calculate scattering amplitudes, it
is necessary to solve the constraint in the following way:
$$\lambda^+=(\dz x^+ +\half(\theta^+\dz \theta^- +\theta^-\dz \theta^+))
e^{h^+}+e^{-h^-},\quad \lambda^-=e^{-h^+},\eqno(2)$$
$$w^+=e^{h^+}[\dz(h^+ +h^-)+x^-
(\dz x^+ +\half(\theta^+\dz \theta^- +\theta^-\dz \theta^+)]+x^- e^{-h^-},
\quad w^-=x^- e^{-h^+} ,$$
where $x^+$ and $x^-$ are the usual longtitudinal bosonic scalars, and
$h^\pm$ are chiral bosons of screening charge $-1$ with the operator-product
$\dz h^+ \dz h^- =z^{-2}$ (note that the relationship between
$w^\pm$, $\lambda^\pm$ and $x^\pm$ closely resembles the
twistor condition of Penrose,$^{14,15}$ $w=x\lambda$).
Because this field redefinition preserves the
operator-product relations of $\lambda^\pm$ and $w^\pm$, the free-field
action of equation 1 is still a free-field action
when $\lambda^\pm$ and $w^\pm$ are replaced by $x^\pm$ and $h^\pm$.
As will be shown later, equation 2 has the effect of
transforming a global constraint on the fields into a constraint on the
U(1) moduli of the surface.$^2$

Using the full SO(9,1) vector, $x^\mu$, one can now construct
N=2 superconformally invariant Lorentz generators out of the free fields.$^3$
This is done by combining the free fields [$x^\mu$, $\Gamma^a$,
$\Gamma^{\bar a}$, $\theta^\pm$, $\varepsilon^\pm$, $h^\pm$] into a
pair of bosonic worldsheet-scalar spacetime-vector superfields, $X^\mu_\pm$
for $\mu=0$ to 9, and a pair of fermionic worldsheet-scalar
spacetime-spinor superfields, $\Theta^\alpha_\pm$ for $\alpha=1$ to 16.
These covariantly transforming superfields satisfy the chirality
conditions, $D_+ X_-^\mu =D_+ \Theta_-^\alpha=
D_- X_+^\mu =D_- \Theta_+^\alpha=0$, and the relation,
$(X_+^\mu -X_-^\mu)\gamma_\mu^{\alpha \beta}=\Theta_+^\alpha \Theta_-^\beta$.
They are not free fields on the worldsheet and are similar in philosophy
to the NSR spin field,$^{16}$
which can only be expressed in terms of free fields
by breaking the manifest Lorentz covariance (note however
that unlike the NSR spin field, they are constructed entirely out of
matter fields). In terms of these covariantly transforming superfields,
the Lorentz generators are $M^{\mu\nu}=\int dz d^2\kappa X^\mu_+
X^\nu_-$, the supersymmetry generators are $S_\alpha=\int dz d^2\kappa
X^\mu_+ \gamma_{\mu\,\alpha\beta}\Theta^\beta_-$, and the massless
vertex operators are $\eta_\mu \int dz d^2\kappa \Theta^\alpha_+
\gamma^\mu_{\alpha\beta}\Theta^\beta_- e^{ik_\nu X^\nu_+}$ for the
vector boson, and $u_\alpha\int dz d^2\kappa \Theta^\alpha_+
e^{ik_\nu X_-^\nu}$ for the spinor fermion.$^3$

The next step is to introduce N=2 super-Virasoro ghosts and a BRST
charge, to bosonize the super-reparametization ghosts, and to
construct picture-changing operators as in the NSR formalism.$^{16}$
Scattering amplitudes are then calculated by evaluating correlation
functions of the vertex operators on N=2 super-Riemann surfaces of
genus $g$, and integrating over the N=2 super-moduli of the surface.
Integration over the fermionic moduli brings down $2(2g-2)$
picture-changing operators whose locations are arbitrary on the surface.
The region of integration for the U(1) moduli, $m_j$ for $j=1$ to $g$,
is the Jacobian variety $C^g/(Z^g+\tau Z^g)$, where
$e^{2\pi i m_j}$ measures the change in phase when $\Gamma^a$ or
$e^{h^+}$ goes around the $j^{th}$ $B$-cycle on the surface.

The only correlation function that is not straightforward to evaluate
is that of the $h^\pm$ fields. Since $e^{h^+ -h^-}$ has negative
conformal weight, it is not possible on a general surface to define
the holomorphic correlation function $<\exp (\sum_k (c_k h^-(z_k)+d_k
h^+(z_k)))>$ without allowing unphysical poles (this situation also
arises with the fields, $\phi$, that come from bosonizing the
bosonic super-reparameterization ghosts, but in that case, the residues
of the unphysical poles are BRST trivial$^{17}$). However on a surface with
the special values of the U(1) moduli, $m_j=\sum_k c_k\int^{z_k}
\omega_j$ where $\omega_j$ is the $j^{th}$ canonical holomorphic
one-form, the unphysical poles are not present. Therefore, it is necessary
for BRST invariance to define the correlation function
$<\exp (\sum_k (c_k h^-(z_k)+d_k
h^+(z_k)))>$ to be proportional to $\prod_{j=1}^g \delta (m_j-\sum_k
c_k\int^{z_k} \omega_j)$, where the proportionality factor is completely
fixed by its conformal properties. In this way, the global constraint on the
$\lambda^\pm$ fields has transformed into
a restriction on the U(1) moduli of the surface.$^2$

After performing the functional integral over the free fields, one is
left with an integrand which depends on the $2(2g-2)$ arbitrary points
where the picture-changing operators have been inserted, and which must
be integrated over the usual $(6g-6+2N)$ Teichmuller parameters and
puncture locations. Although the integrand is not manifestly
Lorentz-invariant, it is possible to use knowledge of the Lorentz
invariance (recall that Lorentz generators have been constructed which
commute with the BRST operator and transform the vertex operators
covariantly) to rewrite the integrand in a manifestly SO(9,1)
invariant form.$^3$

To prove that these Lorentz-invariant expressions are unitary, one
must show agreement with amplitudes obtained using the manifestly
unitary light-cone Green-Schwarz formalism.$^{7-9}$ This has been done
by choosing the moduli (and corresponding Beltrami differentials) for the
surface to be the light-cone interaction points, twists, and internal
$P^+$ momenta, and choosing the locations of the picture-changing
operators to be precisely the interaction points. With this choice for the
surface moduli, it is straightforward to show that the path integrals over
the longtitudinal matter fields [$x^\pm, h^\pm, \theta^\pm, \varepsilon^\pm$]
precisely cancel the path integrals over the ghost fields [$c,b,u,v,\beta^\pm,
\gamma^\pm$], since each bosonic/fermionic longtitudinal
matter field couples in the
same way as a corresponding fermionic/bosonic ghost field.$^4$ The remaining
path integrals over the transverse matter fields give precisely the
light-cone gauge prescription for calculating amplitudes, with the
transverse part of the picture-changing operators becoming the light-cone
Green-Schwarz interaction-point operators.$^1$

Finally, it has been shown$^4$ that these superstring amplitudes (for external
massless boson-boson states in the Type II superstring) are finite
by explicitly checking for divergences when the Riemann surface
degenerates either into one surface with a pinched handle, or into
two surfaces connected by a thin tube (all other possibly divergent regions
in moduli space are related to these by modular transformations).
Note that a similar analysis has not yet been done in the NSR formalism since
before summing over spin structures, the NSR amplitudes are not
divergence-free.$^{18,19}$
When the surface degeneracy corresponds to a pinched handle of radius $R$,
the amplitude behaves like
$(log R)^5 R^{-1}\, dR$ as $R \rightarrow 0$, and is therefore divergence-free.
When the surface degenerates into two
surfaces connected by a tube of radius $R$, the amplitude factorizes into
$A_1 A_2 R^{k^2 -1} \, dR$ as $R \rightarrow 0$, where $k^\mu$ is the momentum
flowing through the tube. If there are vertex operators on both surfaces,
this divergence corresponds to the physical massless pole that is present
in all field theories containing massless particles. In the case when all
vertex operators are located on one of the surfaces, $A_1$ vanishes since
there are no zero modes for the $\theta^\pm$ path integrals (these are
two of the manifest spacetime supersymmetries), and the amplitude is
divergence-free.

It should be noted that
under a change in the locations of the picture-changing operators, the
integrand of the scattering
amplitude changes by a total derivative in the Teichmuller parameters
which can lead to surface term contributions if the
moduli space has a boundary. For example, in the NSR formalism the
need to sum over spin structures before obtaining a finite amplitude
means that a cutoff in the moduli space has to be introduced, which
causes a multiloop ambiguity in the amplitude.$^{20}$
However such an ambiguity does
not occur in the above Green-Schwarz amplitudes since the integrand
is well-behaved when the surface degenerates, and therefore there is
no need to introduce a cutoff and the moduli space contains no boundary.

\vskip 12pt
\centerline{\bf References}

\item{(1)} Berkovits,N., Nucl.Phys.B379 (1992), p.96., hep-th bulletin
board 9201004.

\item{(2)} Berkovits,N., Nucl.Phys.B395 (1993), p.77, hep-th 9208035.

\item{(3)} Berkovits,N., Phys.Lett.B300 (1993), p.53, hep-th 9211025.

\item{(4)} Berkovits,N., Finiteness and Unitarity of Lorentz-Covariant
Green-Schwarz Superstring Amplitudes, King's College preprint KCL-TH-93-6,
March 1993, submitted to Nucl.Phys.B, hep-th 9303122.

\item{(5)} Delduc,F. and Sokatchev,E., Class.Quant.Grav.9 (1992), p.361.

\item{(6)} Berkovits,N., Phys.Lett.B304 (1993), p.249, hep-th 9303025.

\item{(7)} Green,M.B. and Schwarz,J.H., Nucl.Phys.B243 (1984), p.475.

\item{(8)} Mandelstam,S., Prog.Theor.Phys.Suppl.86 (1986), p.163.

\item{(9)} Restuccia,A. and Taylor,J.G., Phys.Rep.174 (1989), p.283.

\item{(10)} Greensite,J. and Klinkhamer,F.R., Nucl.Phys.B291 (1987), p.557.

\item{(11)} Tonin,M., Phys.Lett.B266 (1991), p.312.

\item{(12)} Ademollo,M., Brink,L., D'Adda,A., D'Auria,R.,
Napolitano,E., Sciuto,S., Del Giudice,E., DiVecchia,P., Ferrara,S.,
Gliozzi, F., Musto,R., Pettorini,R., and Schwarz,J., Nucl.Phys.B111
(1976), p.77.

\item{(13)} Ooguri,H. and Vafa,C., Nucl.Phys.B361 (1991), p.469.

\item{(14)} Penrose,R. and MacCallum,M.A.H., Phys.Rep.6C (1972), p.241.

\item{(15)} Sorokin,D.P., Tkach,V.I., Volkov,D.V., and Zheltukhin,A.A.,
Phys.Lett.B216 (1989), p.302.

\item{(16)} Friedan,D., Martinec,E., and Shenker,S., Nucl.Phys.B271
(1986), p.93.

\item{(17)} Verlinde,E. and Verlinde,H., Phys.Lett.B192 (1987), p.95.

\item{(18)} Christofano,G., Musto,R., Nicodemi,F., and
Pettorino,R., Phys.Lett.B217 (1989), p.59.

\item{(19)} Mandelstam,S., Phys.Lett.B277 (1992), p.82.

\item{(20)} Atick,J. and Sen,A., Nucl.Phys.B296 (1988), p.157.

\end